\documentclass[twocolumn,showpacs,preprintnumbers,amsmath,amssymb,aps,superscriptaddress]{revtex4} 
\usepackage{graphicx}
\usepackage{dcolumn}
\usepackage{bm}
\usepackage{mathrsfs}
\usepackage{amssymb}
\usepackage{color}

\newcommand{\bq}{\begin{eqnarray}}
\newcommand{\eq}{\end{eqnarray}}

\renewcommand{\lg}{\langle}
\newcommand{\rg}{\rangle}

\newcommand{\non}{\nonumber}

\begin{document}

%_______________________________________________________________________________________

\title{Superpositions in Atomic Quantum Rings}

\author{R. Kanamoto}

\affiliation{Department of Physics, Meiji University, Kawasaki, Kanagawa 214-8571, Japan}

\author{P. \"{O}hberg}
\affiliation{SUPA, Institute of Photonics and Quantum Sciences, Heriot Watt University, Edinburgh EH14 4AS, United Kingdom}

\author{E. M. Wright}
\affiliation{College of Optical Sciences, University of Arizona, Tucson Arizona 85721, USA}

\date{\today}

%_______________________________________________________________________________________

\begin{abstract}
Ultracold atoms are trapped circumferentially on a ring that is pierced at its center by a flux tube
arising from a light-induced gauge potential due to applied Laguerre-Gaussian fields.
We show that by using optical coherent state superpositions to produce light-induced gauge potentials,
we can create a situation in which the trapped atoms are simultaneously exposed to two distinct flux tubes,
thereby creating superpositions in atomic quantum rings.
We consider the examples of both a ring geometry and harmonic trapping, and in both cases
the ground state of the quantum system is shown to be a superposition of counter-rotating states of the atom trapped on the two distinct flux tubes.
\end{abstract}
\maketitle

%_______________________________________________________________________________________

\section{Introduction}
%_______________________________________________________________________________________

The theoretical concept of creating artificial gauge potentials for neutral atoms
using spatially tailored light fields has now reached maturity 
(see Ref.~\cite{DalGerJuz2011} for a comprehensive theoretical review).
The recent experiments in which an artificial magnetic field was created
in a rubidium Bose-Einstein condensate~\cite{lin_2009b},
has most notably led to the creation of quantized vortices~\cite{LinComJim2009},
spin-orbit coupling~\cite{lin_2011},
and strong effective magnetic fields in optical lattices~\cite{aidelsburger_2011}.
Being able to create artificial gauge fields, both Abelian and
non-Abelian~\cite{osterloh_2005,ruseckas_2005}, has opened up new avenues of research
such as creating exotic matter wave states~\cite{burrello_2010} (quantized vortices
being a simple example~\cite{LinComJim2009}), gauge potentials in optical lattices
allowing for the study of the Harper equation and Hofstadter butterfly~\cite{jaksch_2003},
the creation of Aharanov-Bohm-like effects in atomic gases~\cite{jacob_2007},
and the exploration of analogies and differences with condensed matter and
particle physics ideas~\cite{merkl_2010,lan_2011,mazza_2012}.

Current experiments with artificial gauge potentials rely on a number of different schemes. 
In the continuous gas case Raman transitions have been used in order to tailor the dispersion
relation such that the resulting equation of motion for the center of mass is governed by an 
effective gauge potential \cite{LinComJim2009}. In a similar fashion spin-orbit coupling has also 
been created in a $^{87}$Rb BEC \cite{lin_2011,wang_2012}. Gauge potentials have also been created 
in optical lattice settings where laser-induced tunneling between the adjacent sites is used, 
which can give rise to a nonzero Peierl's phase \cite{aidelsburger_2011}. Similarly the lattice 
configuration also allows for spin-orbit coupling \cite{cheuk_2012}. Nontrivial complex valued tunneling coefficients
can also be created in an optical lattice by mechanically shaking the lattice in a specific manner, and 
thereby obtaining a time averaged tunneling coefficient which can mimic the presence of an effective 
external magnetic field \cite{eckardt_2005,struck_2012,hauke_2012}.

Of particular interest for this paper is the fact that using Laguerre-Gaussian laser fields,
which carry orbital angular momentum, in interactions with a three-level atom,
can yield an induced gauge potential that acts as an effective flux tube~\cite{JRO:JPB:05}.
This is significant in that if the atom is trapped in a radially symmetric potential
and the flux tube pierces the potential, then varying the flux can vary
the angular-momentum properties of the trapped atom~\cite{SFLO:EPL:08,SonFor2009,Ohberg2011}.

The goal of the present paper is to initiate a line of research
involving artificial gauge potentials formed using quantum-mechanical applied light fields,
in particular, optical coherent-state superpositions~\cite{MilWal1994,GlaMac2008}.
The concept is that if the quantum nature of the applied light can be transferred
to the matter waves then we open up the possibility of exposing the atom simultaneously
to a superposition of two or more artificial gauge potentials.
To introduce and explore this possibility, we consider the example of an atomic quantum ring,
both idealized and in a harmonic trap.  An atomic quantum ring involves ultracold atoms
that are trapped circumferentially on a ring that is pierced at its center by a flux tube
arising from the light-induced gauge potential due to
applied Laguerre-Gaussian fields~\cite{SonFor2009,Ohberg2011}.
With a finite flux piercing the ring, the ground state corresponds to a rotating state.
Here we show that by using optical coherent state superpositions to produce 
light-induced gauge potentials,  we can create a situation
in which the trapped atoms are simultaneously exposed to two distinct flux tubes,
thereby creating superpositions in atomic quantum rings.
In particular, we show that the ground state is a quantum superposition of counter-rotating atomic states.
Creation of quantum superposition is also studied for cold atoms in
a rotating ring lattice potential in Refs.~\cite{NAB08,NAB11}.

This paper is organized as follows: In Sec.~\ref{gaugepot} we overview the basic formulation
of the field-induced gauge potential for a single atom
in the electromagnetically induced transparency (EIT) configuration~\cite{DalGerJuz2011,JRO:JPB:05}.
An effective Hamiltonian for the dark-state atom and its eigensolutions
are given for the harmonic and ring potentials.
Section~\ref{optcat} extends the standard dark-state scheme to
the case of a nonclassical control field.
We show that the superposition of distinct motional states has a lower energy
than the statistical mixture of those states.

%_______________________________________________________________________________________

\section{Atomic quantum ring}\label{gaugepot}
%_______________________________________________________________________________________

For the sake of clarity in presentation, and to solidify notation,
in this section we review the basic physical ideas underpinning the atomic quantum ring
for classical applied fields as described by coherent states. 
This idea is applicable both for a single atom and cold atomic ensemble~\cite{aidelsburger_2011} without collisions.
First we set up the model equations as described in Ref.~\cite{JRO:JPB:05},
and then we turn to dark states and the effective gauge potential and flux tube.
Finally we discuss the quantum motional eigenstates
for the cases of a ring geometry and harmonic trapping.

%_______________________________________________________________________________________

\subsection{Model equations}
%_______________________________________________________________________________________

\begin{figure}
\includegraphics[scale=0.5]{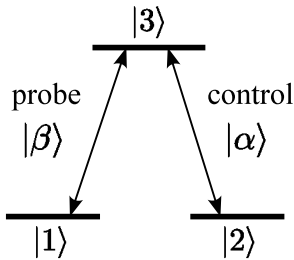}
\caption{
The level scheme. A single atom is irradiated by two lasers:
the probe field that couples $|1\rg$ and $|3\rg$ with the amplitude $\beta$, winding number $\ell_p$, frequency $\omega_p$, and wavenumber $\bm{k}_p$,
and the control field that couples $|2\rg$ and $|3\rg$ with the
amplitude $\alpha$, winding number $\ell_c$, frequency $\omega_c$, and
wavenumber $\bm{k}_c$. Here we take $\ell = \ell_c = - \ell_p$.
}
\label{fig1}
\end{figure}

In a series of recent papers it has been shown how carefully shaped light beams
which are incident on cold atoms can be used for creating strong gauge potentials
in a cloud of neutral atoms.
This effect relies on the interplay between two laser beams and a $\Lambda$-type
level structure of the atoms, as shown in Fig.~\ref{fig1}.
In the following we consider a single atom which is characterized by two hyperfine ground levels
$|1\rg$ and $|2\rg$ of equal energies $\hbar\omega_1=\hbar\omega_2$,
and an electronic excited level $|3\rg$ with energy $\hbar\omega_3$.
The atomic transition could involve, for example, $2\ {}^3S_1$ for the ground Zeeman states of ${}^4$He~\cite{AAKVC88} and $2\ {}^3P_1$ for the excited state.
The atom interacts with two laser beams.
The first beam, which we refer to as the probe beam,
is coupled with the transition $|1\rg \leftrightarrow |3\rg$,
whereas the second beam, the control beam, drives the transition $|2\rg \leftrightarrow |3\rg$.
The probe field is characterized by a wave vector $\bm{k}_p$, and a frequency $\omega_p=ck_p$.
The control laser, on the other hand, has a wave vector $\bm{k}_c$ and a frequency $\omega_c$.
We use Laguerre-Gaussian (LG) beams for the probe and control fields
having the lowest radial quantum number and distinct
orbital angular momenta per photon, which is $\hbar\ell_p$ for the probe field, and 
$\hbar\ell_c$ for the control field, respectively.
Henceforth we choose the control and probe fields to have equal frequency $\omega_p=\omega_c=\omega$, equal magnitude of winding numbers of opposite sign $-\ell_p=\ell_c=\ell$, and take $\bm{k}_c$ and $\bm{k}_p$ collinear along the $z$ axis.

%%%%% H_internal %%%%%

The Hamiltonian for the electronic degree of freedom of an atom interacting with quantized light fields
in the rotating-wave approximation is given by
\bq\label{Hinternal}
\hat{h} (\bm{r})&=&\epsilon_{31}|3\rg \lg 3|\non\\
&-& \hbar \chi(\bm{r})\left(
\hat{a}_p e^{-i\ell \phi}|3\rg \lg 1|
+\hat{a}_c e^{i\ell \phi} |3\rg \lg 2|
+{\rm H.c.}
\right),\non\\
\eq
where $\epsilon_{31}=\hbar (\omega_3-\omega_1-\omega)$
is the energy of the detuning from single-photon resonance,
with $\hbar \omega_j$ being the electronic energy of the atomic level $j=1,2,3$.
Here the Rabi frequency per photon, coupling the ground and excited states,
is denoted as $\chi (\bm{r})$.
The annihilation operators of a photon in the probe and control fields correspond to
$\hat{a}_p$ and $\hat{a}_c$, respectively.
In deriving the above Hamiltonian for the electronic degree of freedom,
we have assumed that the atomic motion is restricted to the two-dimensional plane $\bm{r}=(r,\phi)$
with $r=\sqrt{x^2+y^2}$, due to a tight confinement along the $z$ direction.
The spatial dependence of the Hamiltonian then comes from the space-dependent coupling
associated with the LG beams.

In this section we suppose that the control and probe fields may be described by coherent states
$|\alpha\rg$ and $|\beta\rg$, respectively.
The field state is then expressed by $|\Psi_f \rg = |\alpha\rg_c |\beta\rg_p$
and the effective Hamiltonian for an atom
$\hat{h}_a(\bm{r}) \equiv \lg \Psi_f| \hat{h}(\bm{r}) | \Psi_f \rg$ is given by
\bq
\hat{h}_a(\bm{r})&=&\epsilon_{31}|3 \rg\lg 3|\non\\
&-&\hbar \chi (\bm{r})\left(\beta e^{-i\ell \phi}|3\rg \lg 1|
+\alpha e^{i\ell \phi}|3\rg \lg 2|+{\rm H.c.}
\right). 
\eq
The eigensolutions, arising only from the Hamiltonian of electronic degree of freedom
$\hat{h}_a(\bm{r}) |X\rg = \varepsilon_X |X\rg$,
are denoted by $|D\rg$, $|S\rg$, and $|A\rg$. The state
\bq
|D\rg = \frac{1}{\sqrt{\cal N}}(\alpha e^{i\ell \phi}|1\rg - \beta e^{-i\ell \phi}|2\rg)
\eq
is known as the {\it dark} state~\cite{A:PO:96,H:PT:97,L:RMP:03},
characterized by a zero eigenvalue $\varepsilon_D =0$.
Here ${\cal N} = |\alpha|^2+|\beta|^2$ is the normalization constant.
For $\epsilon_{31}=0$, namely, on resonance, the two other eigensolutions are given by the symmetric and antisymmetric superposition
of the {\it bright} state $|B\rg$ and excited state $|3\rg$,
\bq
&&|S\rg = \frac{1}{\sqrt{2}}(|B\rg + |3\rg),\\
&&|A\rg = \frac{1}{\sqrt{2}}(|B\rg - |3\rg),
\eq
with eigenvalues $\varepsilon_S=|\chi({\bm r})|^2{\cal N}$ and $\varepsilon_A=-|\chi({\bm r})|^2{\cal N}$, respectively, where
\bq
|B\rg = \frac{1}{\sqrt{\cal N}}(\beta^* e^{i\ell \phi}|1\rg + \alpha^* e^{-i\ell \phi}|2\rg).
\eq

If the electronic state of an atom is prepared in the dark state $|D\rg$,
the resonant control and probe beams induce the absorption paths
$|2\rg \to |3\rg$ and $|1\rg \to |3\rg$ which interfere destructively.
This is also the mechanism behind electromagnetically induced transparency~\cite{A:PO:96,H:PT:97,L:RMP:03}.
In such a situation, the transitions to the upper atomic level $|3\rg$ are suppressed,
the atomic level $|3\rg$ is weakly populated, and it is justified to neglect any losses
due to spontaneous emission from the excited state.
We shall hereafter assume that the trapped atom is prepared in the dark state.

%%% H_total %%%

The total Hamiltonian which accounts for both the electronic and motional dynamics is
\bq\label{totalH}
\hat{H} = \frac{\hat{\bm{p}}^2}{2M}+\hat{V}(\bm{r}) + \hat{h}(\bm{r})
\eq
with $M$ being the atomic mass, $\hat{\bm{p}}=-i\hbar \bm{\nabla}$ the momentum operator, and
\bq
\hat{V}(\bm{r}) = V_1(\bm{r})|1\rg\lg 1|+V_2(\bm{r})|2\rg\lg 2|+V_3(\bm{r})|3\rg\lg 3|
\eq
is the trapping potential. The entire quantum state including both the atom and field can be written as
\bq
|\Phi(\bm{r},t)\rg = |\Psi_f\rg\sum_{X=D,S,A}\Psi_X(\bm{r},t)|X\rg, 
\eq
where $\Psi_X(\bm{r},t)$ describes the translational motion of the atom
in one of the three electronic states.
By using this state and the total Hamiltonian~(\ref{totalH}) we arrive at the equation of motion for the three states $\bm{\Psi} = (\Psi_D,\Psi_S,\Psi_A)^T$,
\bq
i\hbar \frac{\partial}{\partial t}\bm{\Psi} = \hat{H}^{\rm (eff)}\bm{\Psi}, 
\eq
where the effective Hamiltonian is given by
\bq\label{Heff}
\hat{H}^{\rm (eff)}= \frac{1}{2M}(i\hbar \nabla - \bm{A})^2 + U
\eq
with
\bq
&&\bm{A}_{X,X'}=-i\hbar \lg X|\nabla X'\rg,\\
&&U_{X,X'}=\varepsilon_{X}\delta_{X,X'} + \lg X|\hat{V}| X'\rg.
\eq
If the internal dynamics is much faster than the external one we can safely assume
the dynamics of the different states to be independent.
In other words, the adiabatic approximation is assumed to hold here.

% ----------------------------
% ----------------------------

%_______________________________________________________________________________________

\subsection{Dark state and the effective flux tube}
%_______________________________________________________________________________________

In the following we assume that the atom remains dominantly in its dark state while moving in space.
The total effective Hamiltonian for the center-of-mass motion of the atom
in the dark state is given from Eq.~(\ref{Heff}) as
\bq
\hat{H}^{\rm (eff)}_{DD}&=&\frac{1}{2M}(i\hbar \nabla - {\bm A}_{DD})^2 + V_{\rm eff}. 
\eq
The resulting gauge potential is defined as
\bq\label{ADD}
{\bm A}_{DD} = -i\hbar \lg D |\nabla D \rg {\bm e}_\phi = \frac{\hbar \ell \sigma}{r}{\bm e}_\phi, 
\eq
where
\bq
\sigma \equiv \frac{|\alpha|^2- |\beta|^2}{{\cal N}}
\eq
is the mean spin of the two LG laser beams.
The effective potential is given by
\bq
V_{\rm eff}=U + \varphi,
\eq
where
\bq
\varphi=\frac{1}{2M}\sum_{X=S,A}\bm{A}_{D,X}\bm{A}_{X,D}.
\eq
For the dark state, the scalar potential is
\bq
\varphi=\frac{\hbar^2}{2M}(\lg D | \nabla D \rg^2 + \lg \nabla D | \nabla D \rg),
\eq
and the effective Hamiltonian for the external motion of an atom in the dark state is
\bq
\hat{H}^{\rm (eff)}_{DD}
&=& \frac{\hbar^2(-\nabla^2 + \lg \nabla D|\nabla D\rg )}{2M}\non\\
&&+\lg D | V(\bm{r}) | D\rg-\frac{\hbar^2}{M}\lg D | \nabla D \rg \nabla  .
\eq
Note that with our Hamiltonian~(\ref{Hinternal}) and LG beams with the lowest radial quantum number,
the dark state depends only on the angle.
In this way, the effective trapping potential $V_{\rm eff}$ is composed of
the external trapping potential and the geometric scalar potential $\varphi$.
Drawing these results together the effective Hamiltonian becomes
\bq
\hat{H}^{\rm (eff)}_{DD} =
-\frac{\hbar^2\nabla^2}{2M}+\frac{\hbar^2 \ell^2}{2Mr^2}+V(\bm{r})
-i \ell \sigma \frac{\hbar^2}{Mr^2} \frac{\partial}{\partial \phi}  .
\eq
For simplicity in notation we hereafter omit the subscript $D$ for the dark state unless otherwise stated.

%%%%% angular momentum %%%%%

The angular-momentum operator for the dark-state atom is given by
\bq
\hat{L}_z = |D\rg \lg D|\left(-i \hbar \frac{\partial}{\partial \phi}\right) |D\rg\lg D|,
\eq
which has an additional term that comes from the dark-state spatial variation,
\bq
\lg D|\left(-i\hbar \frac{\partial}{\partial \phi}\right)|D\rg
= - i\hbar \frac{\partial}{\partial \phi}+r A_{\phi},
\eq
where $A_{\phi} = \hbar \ell \sigma / r$ is the $\phi$ component of the gauge potential Eq.~(\ref{ADD}).
If the mean spin of two lasers $\sigma$ is non-integer,
the orbital angular momentum of the motional state is no longer quantized in the integer units of $\hbar$.
Finally, the effective magnetic flux induced by the effective gauge potential is given by
\bq
\Phi_{\rm mf}=2\pi\hbar\sigma\ell. 
\eq
The effective gauge potential due to the applied LG fields
therefore acts as a flux tube of strength $\Phi_{\rm mf}$.

%_______________________________________________________________________________________

\subsubsection{Ring geometry}\label{ringCS}
%_______________________________________________________________________________________

\begin{figure}
\includegraphics[scale=0.5]{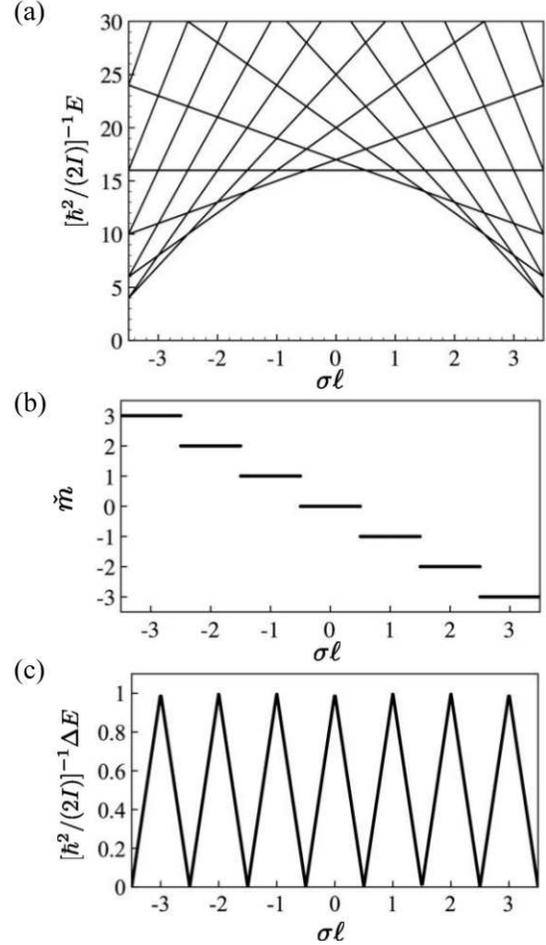}
\caption{
(a) Energy eigenvalues $E_m/(\hbar^2/2I)$, (b) ground-state angular momentum $\check{m}$, and
(c) the lowest excitation energy $\Delta E/(\hbar^2/2I)$, as a function of the mean spin, for $\ell=4$.
}
\label{fig_R}
\end{figure}

First we consider the case that the atom is tightly trapped circumferentially
on a ring of radius $R$ by an external annular potential, which greatly simplifies
the analysis~\cite{SonFor2009,Ohberg2011}.
The effective Hamiltonian at a fixed radius $r=R$ is
\bq\label{H1D}
\hat{H}^{\rm (eff)} = \frac{\hbar^2}{2I}
\left(-\frac{\partial^2}{\partial \phi^2} + \ell^2
- 2i \ell \sigma \frac{\partial}{\partial \phi}\right), 
\eq
where we have defined the rotational inertia $I = MR^2$.
The solutions of the eigenproblem $\hat{H}^{\rm (eff)}\Psi = E \Psi$
for the atomic motional state with the electronic state being the dark state are specified
only by angular-momentum quantum number $m$,
\bq
&& E_m = \frac{\hbar^2}{2I}(\ell^2 + m^2 + 2 \sigma \ell m), \label{ene1D}\\
&& \Psi_m (\phi) = \frac{1}{\sqrt{2\pi}} e^{i m \phi}, \label{wf1D}
\eq
where $m \in \{0,\pm 1,\pm2,\dots\}$.
Generally the energy $E_m$ for a given value of $\sigma \ell$,
and $E_{-m}$ for the value $-\sigma \ell$ are degenerate.
The quantum number for the ground state $\check{m}$ is given by
\bq\label{gsam1D}
\check{m}=-\lfloor\sigma \ell + 1/2\rfloor,
\eq
where $\lfloor s \rfloor$ denotes the floor function applied to the argument $s$.
We note that we have sgn$(\sigma \ell \check{m})<0$. This fact will be important
in the next section in evaluating the energy associated with the superposition.
The energy eigenvalues $E_m$, and the ground-state angular momentum $\check{m}$ for $\ell=4$ are plotted as a function of $\sigma \ell$ in Figs.~\ref{fig_R}(a) and (b), respectively.

Figure~\ref{fig_R}(c) plots the energy separation between the ground and the first excited states,
$\Delta E \equiv E_{\check{m}\pm 1}-E_{\check{m}}= (\hbar^2/2I)\left[\pm 2 (\check{m}+\sigma \ell)-1\right]$.
This energy gap becomes zero at half-integral values of the mean spin $\sigma$,
and the ground-state angular momentum $\check{m}$ changes at these points.
On the other hand, $\Delta E$ takes local maxima at integral values of $\sigma \ell$.
Because of the restriction of the atomic motion to the ring, the excitation energy
as a function of $\sigma \ell$ is independent of $\ell$.

% ----------------------------
% ----------------------------

%_______________________________________________________________________________________

\subsubsection{Harmonic trapping potential}
%_______________________________________________________________________________________

\begin{figure}
\includegraphics[scale=0.5]{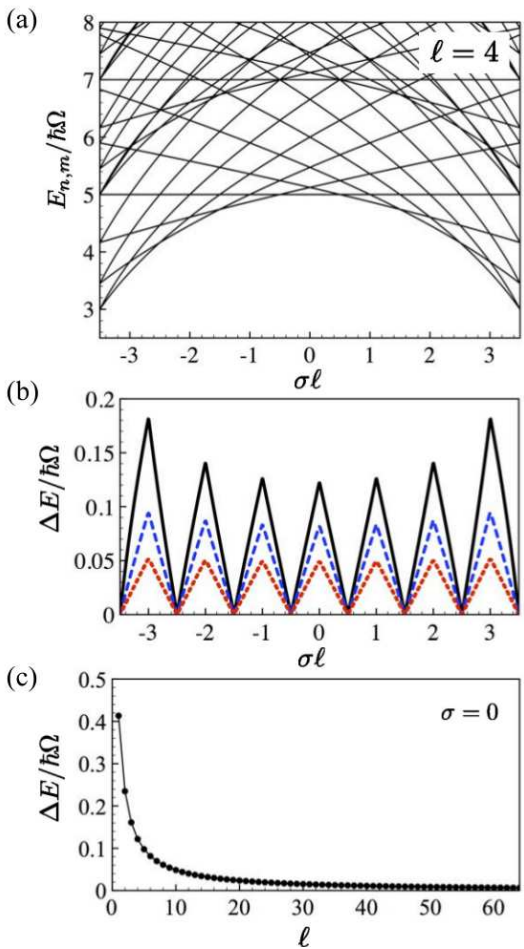}
\caption{
(a) Energy eigenvalues $E_{n,m}/\hbar \Omega$ as a function of $\sigma \ell$. Here we fixed the winding number $\ell=4$.
(b) The lowest excitation energy for $\ell=4$ (line), 6 (dashed), and 10 (dotted).
(c) The lowest excitation energy at $\sigma=0$ as a function of $\ell$.
}
\label{fig_H}
\end{figure}

When the atom is trapped in a harmonic trap  $V(r) = M\Omega^2 r^2/2$,
the eigenproblem for the motional state of the dark-state atom is also analytically solvable~\cite{SFLO:EPL:08}.
With the use of the zero-point oscillator length $r_0=\sqrt{\hbar/(2M\Omega)}$ as the length unit,
the Hamiltonian is
\bq\label{2DH}
\hat{H}^{\rm (eff)}=\hbar \Omega \left[-\nabla^2+\frac{r^2}{4}+\frac{1}{r^2}\left(\ell^2 - 2i\ell \sigma \frac{\partial}{\partial \phi}\right)\right],
\eq
and its eigenvalues and eigenstates are given by
\bq
&&E_{n,m}=\hbar \Omega (2n+\mu_{m} +1), \\
&&\Psi_{n,m}(\bm{r})=\frac{1}{\sqrt{2\pi}}e^{im\phi}f_{n,m}(r), 
\eq
where $n \in \{0,1,\dots\}$ and $m \in \{0,\pm 1, \pm 2,\dots\}$.
The radial dependence of the eigenstate is obtained as
\bq
f_{n,m}(r)= C_{n,m}\left(\frac{r^2}{2}\right)^{\mu_m/2} e^{-r^2/4}
{\cal L}_n^{\mu_m}\!\!\left(\frac{r^2}{2}\right),\label{fnm}
\eq
where $C_{n,m}=\sqrt{n!/\Gamma (n+\mu_m +1)}$ with $\Gamma(x)$ being the gamma function.
The function ${\cal L}_n^{\alpha}(x)$ is the generalized Laguerre polynomial,
parametrized with
\bq
\mu_m = \sqrt{\ell^2 + m^2 + 2\sigma \ell m}.
\eq
Just like the case of the ring geometry, the eigenvalues $E_{n,m}$
for a given value of $\sigma \ell$ are degenerate with $E_{n,-m}$ for the value $-\sigma \ell$.
The quantum numbers $(\check{n},\check{m})$ of ground state are given by
\bq\label{gsam}
\check{n}=0,\quad \check{m}=-\lfloor\sigma \ell +1/2\rfloor,
\eq
and therefore we again have sgn$(\sigma\ell\check{m})<0$ for the ground state.
The energy eigenvalues $E_{n,m}$ for various $n$ with $\ell=4$ are plotted
as a function of $\sigma\ell$ in Fig.~\ref{fig_H}(a).

% ----------------------------
% ----------------------------

%%% Energy separation %%%

Figure~\ref{fig_H}(b) plots the energy separation $\Delta E$ between the ground and the first excited states.
This energy gap similarly becomes zero at half-integral values of the mean spin $\sigma$ and takes local maxima at integral value of $\sigma$.
In contrast to the ring case, for a fixed value of $\ell$ while changing the mean spin $\sigma$, the magnitudes of $\Delta E$ at different integral values of $|\sigma \ell|$
are different values as $\Delta E (\sigma\ell =0) \lesssim \Delta E(|\sigma \ell| = 1) \lesssim \Delta E(|\sigma\ell|=2)\lesssim \cdots$.
When we inspect the energy gap as a function of $\ell$ for a fixed value of mean spin (say, $\sigma=0$), $\Delta E$ is a monotonically decreasing function with respect to $\ell$ as shown in Fig.~\ref{fig_H}(c).

%_______________________________________________________________________________________

\section{Superposition in atomic quantum rings}\label{optcat}
%_______________________________________________________________________________________

The goal of this section is to demonstrate that by making the control field
a quantum superposition of coherent states, the trapped atom can be made to experience
a combination of flux tubes with opposite sign of flux~\cite{SFLO:EPL:08,SonFor2009,Ohberg2011}.
Furthermore we show that the ground state for both harmonic trapping and
a ring geometry is a superposition of rotating atomic states in the individual flux tubes.

\begin{figure}
\includegraphics[scale=0.5]{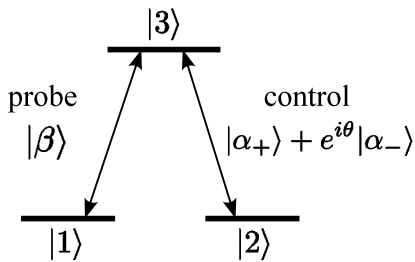}
\caption{
Level scheme to generate flux tubes with opposite sign. The control field is a superposition of coherent states $|\alpha_+\rg$ and $|\alpha_-\rg$.
}
\label{fig4}
\end{figure}

In the following we consider a control field that is described by a quantum superposition of
coherent states $|\alpha_+\rg$ and $|\alpha_-\rg$ (Fig.~\ref{fig4}), whereas the probe field is
described by the single coherent state $|\beta\rg$ as in the previous section.
More specifically we choose $\alpha_{\pm}$ as real and $\beta$ as complex
where a relative phase is included in $\beta$.
The total field state is then written as
\bq
|\Psi_f \rg \propto (|\alpha_+\rg + e^{i\theta}|\alpha_-\rg )|\beta\rg ,
\eq
where $\theta$ is the relative phase between the two coherent state components in the control field.

%_______________________________________________________________________________________

\subsection{Atom-field states for different coherent states}
%_______________________________________________________________________________________

Let us first examine the nature of the atom-field state corresponding to each coherent-state component of the quantum superposition separately.
For the respective component associated with either of coherent states $|\alpha_\pm \rg$,
the total quantum state, including field, and the atomic motional and internal states may be written as
\bq
|\Phi_{\pm}(\bm{r}) \rg = \Psi_{\pm}(\bm{r})|D_{\pm}(\phi)\rg|\alpha_{\pm}\rg |\beta\rg,
\eq
where the dark state in each component is respectively given from the discussion in the previous section:
\bq
|D_{\pm}(\phi)\rg = \frac{1}{\sqrt{{\cal N}_{\pm}}}
\left(\alpha_{\pm} e^{i\ell \phi}|1\rg -\beta e^{-i\ell\phi}|2\rg\right) ,
\eq
with ${\cal N}_{\pm}=\alpha_{\pm}^2+|\beta|^2$ a normalization constant.
We also define the mean spin of each component as
\bq
\sigma_{\pm}\equiv \frac{\alpha_{\pm}^2-|\beta|^2 }{\alpha_{\pm}^2+|\beta|^2}.
\label{sigmapm}
\eq

Now we describe our scheme more specifically: in particular, we want to choose the amplitudes $\alpha_\pm$ such that the two coherent-state components correspond to flux tubes of opposite sign which requires that the mean spins of the two components are opposite in sign $\sigma_+=-\sigma_-=\sigma$.  Using Eq.~(\ref{sigmapm}) we find that the coherent-state amplitudes have to obey either of the following two conditions
\bq\label{i}
{\rm (i)}\quad |\beta|^2 = \alpha_+ \alpha_-
\eq
or
\bq\label{ii}
{\rm (ii)}\quad |\beta|^2 = - \alpha_+ \alpha_-.
\eq
When either of these two conditions is satisfied, the two components of the coherent state
correspond to situations in which the trapped particle will experience flux tubes with fluxes
$\Phi_{\rm mf}=\pm 2\pi\hbar|\sigma\ell|$ of equal magnitude but opposite sign.

From the perspective of the fragility of optical coherent-state superpositions
against interactions with their environment, case (i) above is preferable and we hereafter focus our 
attention on this case.  This follows since the optical coherent-state superpositions decay as
$\exp(-|\alpha_+ -\alpha_-|^2\gamma t/2)$~\cite{MilWal1994}, with $\gamma$ a constant
dependent on the specific dissipation mechanism.
Glancy and Macedo de Vasconcelos~\cite{GlaMac2008} have reviewed methods for producing 
optical coherent-state superpositions.
For our present purposes we require a superposition of coherent states
that are macroscopically {\it distinguishable} but not necessarily macroscopically separated,
with the mean photon numbers $|\alpha_\pm|^2$ separated by only a few quanta.
The feasibility of creating such optical coherent state superpositions was already alluded to
in the seminal work of Ref.~\cite{SonCavYur1990}.

% ----------------------------
% ----------------------------
%_______________________________________________________________________________________

\subsection{Atom-field state for coherent-state superposition}
%_______________________________________________________________________________________

We next examine the normalized quantum state for the combined atom-field system
including both components of the coherent state
\bq\label{swf}
|\Phi (\bm{r},t)\rg = \frac{|\Phi_+\rg + e^{i\theta}|\Phi_-\rg}
{\sqrt{2[1+|\lg \Phi_+|\Phi_- \rg |\cos(\theta+\psi)]}},
\eq
where $\lg \Phi_+|\Phi_- \rg \equiv|\lg \Phi_+|\Phi_- \rg| e^{i\psi} $.
We note that the relative phase between the field coherent-state components
also appears in the atom-field quantum state.
Consistency demands that the normalized state vector (\ref{swf}) obeys the Schr\"odinger equation
\bq\label{Scheq}
i\hbar \frac{\partial}{\partial t}|\Phi(\bm{r},t)\rg = \hat{H} |\Phi(\bm{r},t)\rg, 
\eq
where $\hat{H}$ is given by Eq.~(\ref{totalH}).

In choosing the form of the quantum state in Eq.~(\ref{swf}) we have tacitly assumed that it contains only the consistent dark-state components
$|\beta\rg|\alpha_{\pm}\rg|D_{\pm}\rg$ of the Hamiltonian $\hat{h}_a (\bm{r})$ that are immune to decay from the excited state. 
This choice of the form of the quantum state is motivated by the common notion that non-dark state components such as 
$|\beta\rg|\alpha_{\mp}\rg|D_{\pm}\rg$, for which the field and atomic states  are incompatible for a dark state,
will decay due to spontaneous emission from the excited state.  For example, one might venture an ansatz for the initial combined atom-field state as 
$|\Psi_f\rg (\Psi_+(\bm{r})|D_+\rg+\Psi_-(\bm{r})|D_-\rg)$,  but following the decay of the non-dark-state components in this state vector leads to the quantum state
in Eq. (\ref{swf}) to within a normalization constant.  In this connection we note that for case (i) above the two coherent states $|\alpha_\pm\rangle$, though not macroscopically separated, are macroscopically distinguishable, 
meaning that $|\langle\alpha_-|\alpha_+\rangle| \ll 1$.  The macroscopic distinguishability of the two coherent states is required to substantiate the claim of distinct decay properties for the dark and non-dark states above, and the requirement that they not be macroscopically separated is based on wanting to minimize the detrimental effects of decoherence on the field superposition.
Finally, we have assumed that any back action of the single atom back on the field is neglected: This is valid as long as the large amplitude $|\alpha_{\pm}|$ of the coherent state ensures that the mean photon number is much larger than unity. This is our case, because we only demand that $|\alpha_+-\alpha_-|$ be small, but $|\alpha_{\pm}|$ can be arbitrarily large. In such a case, any back action of atoms onto the large amplitude field will be small.  We note, however, that the situation is much different in a cavity, where the mean number of photons is significantly restricted.

Our next goal is to derive equations of motion for the atomic motional wave functions $\Psi_\pm$
corresponding to the two coherent-state components.
However, this is complicated by the fact that $|\Phi_\pm\rangle$ need not be orthogonal
which originates from the fact that the coherent states $|\alpha_\pm \rg$ are not orthogonal.
This means that cross terms between the components must be retained.
In particular, we need the matrix elements of the Hamiltonian
with respect to the dark states $|D_\pm\rangle$, which are given as
\bq
&&\lg D_{\pm}|H|D_{\pm}\rg
= \hat{K}-\frac{\hbar^2}{M}\lg D_{\pm} | \nabla D_{\pm} \rg \nabla, \\
&&\lg D_{\pm}| H |D_{\mp} \rg = \hat{K} \lg D_{\pm}|D_{\mp} \rg
-\frac{\hbar^2}{M}\lg D_{\pm}|\nabla D_{\mp} \rg\nabla, \label{ODH}
\eq
where
\bq
\hat{K}\equiv \frac{\hbar^2}{2M}\left(-\nabla^2 + \frac{M^2 \Omega^2 r^2}{\hbar^2} + \frac{\ell^2}{r^2}\right)
\eq
for the harmonic potential, and
\bq
\hat{K} \equiv \frac{\hbar^2}{2I}\left(-\frac{\partial^2}{\partial \phi^2} + \ell^2\right)
\eq
for the ring trap of radius $R$. The cross terms between different dark states are calculated as
\bq
&&\lg D_{\pm}|D_{\mp}\rg=\frac{\alpha_+\alpha_-+|\beta|^2}{\sqrt{{\cal N}_+{\cal N}_-}},\\
&&\lg D_{\pm}| \nabla D_{\mp} \rg = \frac{i \ell (\alpha_+\alpha_--|\beta|^2)}{r\sqrt{{\cal N}_+{\cal N}_-}}.
\eq
The above results will be used to obtain the equations of motion for the wavefunctions $\Psi_\pm$ of the two coherent-state components by substituting the state vector in Eq. (\ref{swf}) into Eq. (\ref{Scheq}), and projecting onto the two (non-orthogonal) components.
In the following we deal with the two conditions set out in Eqs. (\ref{i}) and (\ref{ii}) separately.

%_______________________________________________________________________________________

\subsubsection{Case $ (i) \ |\beta|^2 = \alpha_+ \alpha_-$}
%_______________________________________________________________________________________

\begin{figure}
\includegraphics[scale=0.6]{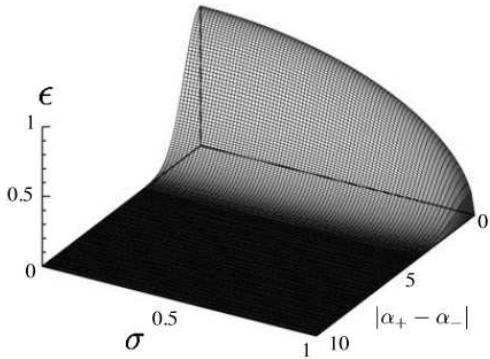}
\caption{
Parameter $\epsilon$ as functions of $|\alpha_+-\alpha_-|$ and $\sigma$.
}
\label{fig_eps}
\end{figure}

For this case the cross terms between the dark states reduce to
\bq
\lg D_{\pm}|D_{\mp}\rg = \sqrt{1-\sigma^2},\quad\lg D_{\pm}|\nabla D_{\mp} \rg=0.
\eq
Then projection of the Schr\"odinger equation~(\ref{Scheq}),
multiplying $\lg D_{\pm}| \lg \alpha_{\pm} | \lg \beta|$ from the left,
generates the following set of equations for the wavefunctions $\Psi_{\pm}(\bm{r})$
for the atomic external degree of freedom
\begin{widetext}
\bq
&&i\hbar \frac{\partial}{\partial t}[\Psi_+ + \epsilon e^{i\theta} \Psi_-]
=\left(\hat{K}-\frac{\hbar^2}{M}\lg D_+|\nabla D_+\rg\nabla\right)\Psi_++\epsilon e^{i\theta}\hat{K}\Psi_-, \label{eqfora}\\
&&i\hbar \frac{\partial}{\partial t}[\epsilon e^{-i\theta} \Psi_++\Psi_-]
=\epsilon e^{-i\theta}\hat{K}\Psi_+ + \left(\hat{K}-\frac{\hbar^2}{M}\lg D_-|\nabla D_- \rg \nabla\right)\Psi_- ,\label{eqforb}
\eq
\end{widetext}
where the following real parameter characterizes the non-orthogonality of two components
\bq\label{paraeps}
\epsilon = \lg \alpha_+|\alpha_- \rg\sqrt{1-\sigma^2}=e^{-|\alpha_+-\alpha_-|^2}\sqrt{1-\sigma^2}.
\eq
Figure~\ref{fig_eps} shows the dependence of $\epsilon$ on $\sigma$ and $|\alpha_+-\alpha_-|$, and shows that we may control the size of $\epsilon$ by controlling the difference between the coherent-state amplitudes $\alpha_\pm$.  In keeping with case (i) reflected in Eq. (\ref{i}), if $\alpha_\pm$ have the same sign and differ in magnitude squared by a few quanta we may control $0\le\epsilon\le 1$ for a given $\sigma$.  Typically we want $\epsilon$ small, say $1/10$, but not too small.

% ----------------------------
% ----------------------------

First we consider the ring geometry as this allows us to illustrate the basic ideas involved with the least complexity.  For the ring case the atom is constrained to move on a circle of radius $R$ with position parametrized by the
azimuthal angle $\phi$. The atomic ring radius $R$ is typically a few micrometers, which is smaller than
the typical beam waist $w_0 \sim 100 \ \mu$m of the LG beams.
In order to evaluate the energy of superposition state in unit of $\hbar^2/(2I)$, we use the ansatz for the ground-state wavefunctions
\bq\label{wf1d}
\Psi_{\pm} (\phi) \propto (\xi_{\pm} e^{i\check{m}\phi} + \zeta_{\pm} e^{-i\check{m}\phi})e^{-iEt/\hbar}  ,
\eq
where $\xi_{\pm}, \zeta_{\pm}$ are $c$ numbers.  This ansatz is motivated by the fact that in the approximation that the coherent-state components are treated as orthogonal $(\epsilon\rightarrow 0)$, the solutions of Eqs. (\ref{eqfora}) and (\ref{eqforb}) should coincide with those given 
in Sec.~\ref{ringCS}.  In particular, the solutions  $\Psi_+ \propto e^{i\check{m}\phi}$ and $\Psi_- \propto e^{-i\check{m}\phi}$ correspond to
the rotating ground-state eigenfunctions of the Hamiltonian~(\ref{H1D}) for $+\sigma\ell$ and $-\sigma\ell$.  Furthermore, the energies of the rotating eigenfunctions $\Psi_{\pm}$ are degenerate,
\bq
E_0=\frac{\hbar^2}{2I}(\ell^2 + \check{m}^2 + 2 \sigma \ell \check{m}).
\eq
Figure \ref{fig_R} illustrates the degeneracy of the ground states for values $\pm\sigma\ell$ for the case with no cross coupling $\epsilon=0$.  However, in the presence of cross coupling the angular momentum states with $\pm\check{m}$ become intermixed, hence the form of the ansatz (\ref{wf1d}).  The key question to be addressed is whether cross coupling with $\epsilon\ne 0$ can lower the ground-state energy of the system. If so, then the state vector in Eq. (\ref{swf}), which represents a superposition of the atom trapped simultaneously on the two different flux tubes, will have an energy lower than a simple mixture state of the atom trapped on one or the other of the two flux tubes that has energy $E_0$. Furthermore, the energetically favored ground state will have the form of a quantum superposition of the atom in the counter-rotating angular momentum states $\pm\hbar\check{m}$.

To determine the ground-state energy in the presence of cross coupling, we substitute~(\ref{wf1d}) into Eqs.~(\ref{eqfora}) and (\ref{eqforb}), and
use the orthogonality of the spatial modes $e^{\pm i\check{m}\phi}$ to obtain equations for
$\xi_{\pm}$ and $\zeta_{\pm}$ as
\begin{widetext}
\bq
&&\frac{\hbar^2}{2I}\left[
\begin{array}{cc}
\ell^2 + \check{m}^2 +2\sigma \ell \check{m}& e^{i\theta}\epsilon (\ell^2+ \check{m}^2)\\
e^{-i\theta}\epsilon (\ell^2 + \check{m}^2) & \ell^2+\check{m}^2-2\sigma \ell \check{m}
\end{array}
\right]
\left[
\begin{array}{c}
\xi_+\\
\xi_-
\end{array}
\right]
=E\bm{A}
\left[
\begin{array}{c}
\xi_+\\
\xi_-
\end{array}
\right], \\
&&\frac{\hbar^2}{2I}\left[
\begin{array}{cc}
\ell^2 + \check{m}^2 -2\sigma \ell \check{m}& e^{i\theta}\epsilon (\ell^2+ \check{m}^2)\\
e^{-i\theta}\epsilon (\ell^2 + \check{m}^2) & \ell^2+\check{m}^2+2\sigma \ell \check{m}
\end{array}
\right]
\left[
\begin{array}{c}
\zeta_+\\
\zeta_-
\end{array}
\right]
=E\bm{A}
\left[
\begin{array}{c}
\zeta_+\\
\zeta_-
\end{array}
\right], 
\eq
\end{widetext}
where
\bq\label{matA}
\bm{A}=\left[
\begin{array}{cc}
1 & \epsilon e^{i\theta}\\
\epsilon e^{-i\theta} & 1
\end{array}
\right].
\eq
These equations have common eigenvalues $E=\{E_+,E_-\}$,
\bq
\left(\frac{\hbar^2}{2I}\right)^{-1}\!\!\! E_{\pm}=\ell^2+\check{m}^2 \pm \frac{2\sigma \ell \check{m}}{\sqrt{1-\epsilon^2}}.
\eq
Among these two eigenvalues, only $E_+$ is relevant here as it coincides with the degenerate ground-state
energy $E_0$ in the limit $\epsilon \to 0$.
The energy difference associated with the superposition $\delta E \equiv E_+-E_0$ is therefore given by
\bq
\left(\frac{\hbar^2}{2I}\right)^{-1}\!\!\! \delta E = 2 \sigma \ell \check{m} \left(\frac{1}{\sqrt{1-\epsilon^2}}-1\right), 
\eq
which is {\it negative} since sgn$(\sigma \ell \check{m})<0$ for the ground state [see the discussion surrounding Eq.~(\ref{gsam1D})].
The superposition state thus has a lower energy than that of the mixed states of two coherent-state components.
For small $\epsilon$ this reduction in energy is written as
$(\hbar^2/2I)^{-1} \delta E\simeq - |\sigma \ell \check{m}| \epsilon^2$, which scales as $\epsilon^2$.
However, we note that the order of magnitude $[\hbar^2/(2I)]^{-1}\delta E$ can be much larger than $\epsilon^2$ because of the prefactor $|\sigma \ell \check{m}|$. Remembering that the ground-state angular momentum is given by Eq. (\ref{gsam1D}), and that we can adjust so that $\sigma \ell$ is an integer, the prefactor $|\sigma \ell \check{m}|$ is a square of an arbitrary integer. Furthermore, as one can increase the OAM $\ell$~\cite{Zei12} and entangle OAM states in high dimensions~\cite{Dada11}, the energy gap can be as large as $\hbar^2/2I$ even if $\epsilon^2$ is small.

\begin{figure}
\includegraphics[scale=0.48]{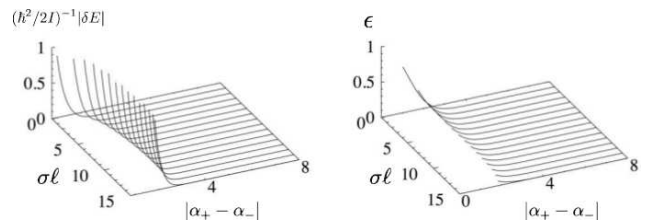}
\caption{
Magnitude of energy reduction as functions of the integer values of $\sigma \ell$ and $|\alpha_+-\alpha_-|$
for $\ell=16$.
}
\label{fig_1D_sup}
\end{figure}

It is preferable to have a larger reduction in energy $|\delta E|$ in terms of robustness.
On the other hand, $|\delta E|$ should not be larger than the lowest excitation energy
$\Delta E$ from the ground state in the absence of superposition;
otherwise the ansatz~(\ref{wf1d}) is no longer valid.
Therefore, from an examination of the eigenvalue structure, we employ an integral value of $\sigma \ell$,
where the energy gap takes maximum value $\Delta E=\hbar^2/(2I)$ .
Figure~\ref{fig_1D_sup} plots the magnitude of the energy gain $|\delta E|$ in the region where $|\delta E| < \Delta E$. The corresponding parameter $\epsilon$ is also plotted.
When $|\delta E|/(\hbar^2/2I)$ is neither too small ($\sim 0$) nor too large ($\sim 1$), e.g., at $|\alpha_+-\alpha_-| \simeq 3$, the superposition is feasible.

The eigenvectors $\{\xi_{\pm}, \zeta_{\pm}\}$, corresponding to the eigenvalue $E_+$, give the admixture of the distinct rotational states with winding numbers $\pm\check{m}$ in the ground-state superposition. These eigenvectors are obtained as
\bq
&&{}^t[\xi_+,\xi_-] \propto {}^t[-\epsilon e^{i\theta},1-\sqrt{1-\epsilon^2}],\\
&&{}^t[\zeta_+,\zeta_-] \propto {}^t[-\epsilon e^{i\theta},1+\sqrt{1-\epsilon^2}].\label{ieigenvec}
\eq
Thus for $\epsilon \ll 1$ we find
\bq
\left | {\xi_-\over\xi_+}\right |^2 = \left | {\zeta_+\over\zeta_-}\right |^2 = {\epsilon^2\over 4} \ll 1  ,
\eq
which means that there is little mixing between the rotational states, and we have a superposition of counter-rotating states to a high degree.
Figure~\ref{fig_1D_sup} plots the magnitude of the energy reduction and the real parameter $\epsilon$
only in the region where $|\delta E|$ is smaller than the energy gap represented in Fig.~\ref{fig_H}.

% ----------------------------
% ----------------------------

We now repeat the same procedure for the case of harmonic trapping which modifies the details but not the concept of the superposition state.
We employ a similar ansatz for wave functions,
\bq\label{wf2d}
\Psi_{\pm}(r,\phi) \propto [\xi_{\pm} e^{i\check{m}\phi}+\zeta_{\pm}e^{-i\check{m}\phi} ]R_{\check{m}}(r) e^{-iEt/\hbar}, 
\eq
where
\bq
R_{\check{m}}(r)&\equiv& f_{0{\check{m}}}(r)\non\\
&=&\sqrt{\frac{2M\Omega}{\hbar \Gamma(\mu_{\check{m}}+1)}}
\left(\frac{M\Omega r^2}{\hbar}\right)^{\mu_{\check{m}}/2}e^{-(M\Omega/2\hbar)r^2}\non\\
\eq
is the radial eigenfunction of Eq.~(\ref{fnm}) for the ground-state quantum numbers $n=0, m=\check{m}$.
Similar calculation as in the ring-geometry case yields equations for $\xi_{\pm}, \zeta_{\pm}$ for the harmonic-trapping case as
\bq
&&\hbar \Omega\left[
\begin{array}{cc}
\eta+\sigma \ell \check{m}/\mu_{\check{m}} & e^{i\theta}\epsilon \eta\\
e^{-i\theta} \epsilon \eta & \eta-\sigma \ell \check{m}/\mu_{\check{m}}
\end{array}
\right]
\left[
\begin{array}{c}
\xi_+\\
\xi_-
\end{array}
\right]
=E\bm{A}
\left[
\begin{array}{c}
\xi_+\\
\xi_-
\end{array}
\right], \non\\
&&\hbar\Omega\left[
\begin{array}{cc}
\eta - \sigma \ell \check{m}/\mu_{\check{m}} & e^{i\theta}\epsilon \eta\\
e^{-i\theta}\epsilon \eta & \eta + \sigma \ell \check{m}/\mu_{\check{m}}
\end{array}
\right]
\left[
\begin{array}{c}
\zeta_+\\
\zeta_-
\end{array}
\right]
=E\bm{A}
\left[
\begin{array}{c}
\zeta_+\\
\zeta_-
\end{array}
\right], \non\\
\eq
where $\bm{A}$ is given by Eq.~(\ref{matA}), and
$\eta \equiv \mu_{\check{m}}+1 -\sigma\ell \check{m}/\mu_{\check{m}}$.
The common eigenvalues are
\bq
\frac{E_{\pm}}{\hbar \Omega}
=\eta
\pm \frac{\sigma \ell \check{m}}{\mu_{\check{m}} \sqrt{1-\epsilon^2}}, 
\eq
and again $E_+$ is relevant, since in the limit $\epsilon\to 0$ it coincides with
the degenerate ground-state energy in the harmonic-trapping potential 
$E_0  = \hbar \Omega (\mu_{\check{m}}+1)$.
The energy difference between the superposition and statistical mixture is
\bq
\frac{\delta E}{\hbar \Omega} = \frac{\sigma \ell \check{m}}{\mu_{\check{m}}}\left(\frac{1}{\sqrt{1-\epsilon^2}}-1\right),
\eq
which is negative by virtue of the fact that sgn$(\sigma\ell\check{m})<0$, i.e., the superposition has a lower energy than the mixture.
For small $\epsilon$, this reduction in energy is written as
$\delta E /\hbar \Omega \simeq -|\sigma \ell \check{m}| \epsilon^2 / (2\mu_{\check{m}})$, which again scales as $\epsilon^2$ multiplied by the prefactor $\sigma \ell \check{m}/\mu_{\check{m}}$. In the harmonic-trapping case, $\mu_{\check{m}}$, which increases for a larger $\ell$, works to decrease the energy gap $\delta E$. Because of this factor, the ring case is more preferable than the harmonic-trapping case to a have more robust ground state. As the oscillator frequency, we employ $\Omega \simeq 2\pi \times 40$ s$^{-1}$, and the corresponding radius of the atomic cloud would be $20\ \mu$m. With this frequency $\Omega$, we note that the energy unit $\hbar \Omega$ is the same order of magnitude of the previous ring-trap case.
Thus, we have a superposition of counter-rotating states as a ground state in the case of the harmonic trapping, too.

%_______________________________________________________________________________________

\subsubsection{Case $(ii)\ |\beta|^2 = -\alpha_+ \alpha_-$}
%_______________________________________________________________________________________

In this case the cross terms of dark states reduce to
\bq
\lg D_{\pm} | D_{\mp}\rg = 0, \quad
\lg D_{\pm} | \nabla D_{\mp} \rg = -\frac{i\ell}{r}\sqrt{1-\sigma^2}.
\eq
Following the same procedure as in case (i), the Schr\"odinger equation is obtained as
\begin{widetext}
\bq
&&i\hbar \frac{\partial}{\partial t}\Psi_+(\bm{r})
=\left(\hat{K}-\frac{\hbar^2}{M}\lg D_+ | \nabla D_+ \rg \nabla\right)\Psi_+(\bm{r})
+ \epsilon e^{i\theta} \frac{i \hbar^2 \ell}{Mr}\nabla \Psi_-(\bm{r}), \\
&&i\hbar \frac{\partial}{\partial t}\Psi_-(\bm{r})
=\epsilon e^{-i\theta}\frac{i\hbar^2 \ell}{Mr}\nabla \Psi_+(\bm{r})
+\left(\hat{K}-\frac{\hbar^2}{M}\lg D_- | \nabla D_- \rg \nabla\right)\Psi_-(\bm{r}), 
\eq
\end{widetext}
where $\epsilon$ is defined by Eq.~(\ref{paraeps}).
We again study the cases of ring potential, and harmonic potential, respectively, and
show only the results here without commentary.

With the use of the same ansatz~(\ref{wf1d}),
we obtain the difference in the energy of superposition and that of mixture $\delta E \equiv E_+ - E_0$ as
\bq
\left(\frac{\hbar^2}{2I}\right)^{-1}\!\!\! \delta E =2\ell \check{m}  (\sqrt{\epsilon^2 + \sigma^2}-\sigma), 
\eq
where we again employed the solution that coincides with $E_0$ in the limit $\epsilon \to 0$.
For small $\epsilon$, this is expanded as
\bq
\left(\frac{\hbar^2}{2I}\right)^{-1}\!\!\!\delta E \simeq \frac{\ell \check{m} \epsilon^2}{\sigma}, 
\eq
which is again negative, meaning that the superposition state is energetically favored.

For the integer values of $\sigma \ell$, the condition $|\delta E| < \Delta E$ turned out to be
identical to the case (i).

Eigenvectors for the ground state $E_+$ are
\bq
&&{}^t[\xi_+,\xi_-] = {}^t[\epsilon e^{i\theta},\sigma-\sqrt{\epsilon^2 + \sigma^2}],\\
&&{}^t[\zeta_+,\zeta_-] = {}^t[\epsilon e^{i\theta},\sigma+\sqrt{\epsilon^2 + \sigma^2}],\label{iieigenvec}
\eq
and for $\epsilon \ll 1$ we have
\bq
\left|\frac{\xi_-}{\xi_+}\right|^2 = \left|\frac{\zeta_+}{\zeta_-}\right|^2=\frac{\epsilon^2}{4\sigma^2} \ll 1.
\eq
This result again means that there is little mixing between the rotational states,
and we have a superposition of counter-rotating states to a high degree.

The ansatz~(\ref{wf2d}) leads to the energy difference 
\bq
\frac{\delta E}{\hbar \Omega} = \frac{\ell \check{m}}{\mu_{\check{m}}} (\sqrt{\epsilon^2 + \sigma^2} -\sigma), 
\eq
and the corresponding eigenvectors are given by Eq~(\ref{iieigenvec}).
For $\epsilon \ll 1$,
\bq
\frac{\delta E}{\hbar \Omega} = \frac{\ell \check{m}}{2\sigma \mu_{\check{m}}}\epsilon^2.
\eq
The energy associated with the superposition and the mixing rate of the rotational states are the order of $\epsilon^2$.

%_______________________________________________________________________________________

\section{Conclusion}\label{conclusion}

In summary, we have introduced the idea of using quantized light fields
for the creation of artificial gauge fields, and shown that it can yield superpositions
in atomic quantum rings. The underlying concept is that by using superpositions of optical coherent states,
one can expose an atom simultaneously to a combination of artificial gauge fields,
or in our specific example, to a combination of flux tubes.
For the atomic quantum ring this was shown to lead to a ground state that was
a superposition of counter-rotating atomic states.

It should be noted that a superposition of counter-rotating atomic states can also be created using synthetic spin-orbit coupling~\cite{lin_2011,jacob_2007,SonFor2009}. The gauge potential stems in this case from classical light fields and is also static, where each component of the resulting atomic pseudo spin can experience opposite constant magnetic fields. Artificial gauge potentials formed using quantum mechanical applied light fields, with the possibility of exposing the atom simultaneously to a superposition of two or more artificial gauge potentials, offers some intriguing concepts. Not only does it provide a route towards mesoscopic superposition states of quantum gases, 
but it may also allow for creation of entanglement between light fields and motional degrees of freedom in the quantum gas. In addition it may  provide a route to construct a back action between the gauge field and the atomic center-of-mass state by relying on strong coupling between the constituents, and by doing so simulate a dynamical gauge theory. From a quantum simulator point of view this would be important, as it would open up the possibility to emulate field theories known from particle physics and the standard model.

It is certainly tempting to extend these ideas in several ways including inclusion of many-body effects, treatment of more general quantized light fields, using squeezed light field for the pump and for the probe fields, coupling between the light and matter-wave fields in an optical cavity, and the application to more general geometries such as atomic motion in a combination of gauge fields of induced optical lattices.

P.\"O. acknowledges support from the UK EPSRC, and R.K. acknowledges support by Grant-in-Aid for Scientific Research from MEXT (Grant No. 23104712) Japan.

\ \ \\

\ \ \\

%_______________________________________________________________________________________

%_______________________________________________________________________________________

%_______________________________________________________________________________________

\end{document}